# Single rare-earth ions as atomic-scale probes in ultra-scaled transistors

Qi Zhang[1,2,5,6]†, Guangchong Hu[2]†, Gabriele G. de Boo[2], Milos Rancic[3], Brett C. Johnson[4], Jeffrey C. McCallum[4], Jiangfeng Du[1,5,6], Matthew J. Sellars[3], Chunming Yin[2]* & Sven Rogge[2]

[1] CAS Key Laboratory of Microscale Magnetic Resonance and Department of Modern Physics, University of Science and Technology of China, Hefei 230026, China
[2] Centre of Excellence for Quantum Computation and Communication Technology, School of Physics, University of New South Wales, Sydney, New South Wales 2052, Australia.
[3] Centre of Excellence for Quantum Computation and Communication Technology, RSPE, Australian National University, Canberra, Australian Capital Territory 0200, Australia.
[4] Centre of Excellence for Quantum Computation and Communication Technology, School of Physics, University of Melbourne, Melbourne, Victoria 3010, Australia
[5] Hefei National Laboratory for Physical Sciences at the Microscale, University of Science and Technology of China, Hefei 230026, China
[6] Synergetic Innovation Center of Quantum Information and Quantum Physics, University of Science and Technology of China, Hefei 230026, China
†These authors contributed equally to this work.
*E-mail: c.yin@unsw.edu.au

**Abstract**
**Continued dimensional scaling of semiconductor devices has driven information technology into vastly diverse applications. As the size of devices approaches fundamental limits, metrology techniques with nanometre resolution and three-dimensional (3D) capabilities are desired for device optimisation. For example, the performance of an ultra-scaled transistor can be strongly influenced by the local electric field and strain. Here we study the spectral response of single erbium ions to applied electric field and strain in a silicon ultra-scaled transistor. Stark shifts induced by both the overall electric field and the local charge environment are observed. Further, changes in strain smaller than $3 \times 10^{-6}$ are detected, which is around two orders of magnitude more sensitive than the standard techniques used in the semiconductor industry. These results open new possibilities for non-destructive 3D mapping of the local strain and electric field in the channel of ultra-scaled transistors, using the single erbium ions as ultra-sensitive atomic probes.**

Complementary metal-oxide-semiconductor (CMOS) has been the most widely used technology in very-large-scale integration for decades due to its high noise immunity and low static power consumption. As integrated circuits scale towards the sub-10-nm nodes and clock speeds reach above 3-4 GHz, heating becomes a significant limit for traditional CMOS devices[1]. Thus, a high carrier mobility, which is strongly influenced by the local environment inside the channel, is becoming a crucial requirement for improving transistor performance. Recently in the semiconductor industry, there has been a departure from the traditional planar device geometry[2]. The multilayer complexity has made it necessary to precisely control the electric field and strain in the device channel as they can strongly affect the carrier mobility[3]. Yet a major problem arises: as the device size shrinks further, process variations[4] make it increasingly difficult to predict transistor properties only with finite element simulations. Therefore, non-destructive 3D strain and electric field mapping in the channel of ultra-scaled transistors would be a leap forward in understanding and optimising these devices[5]. However, this is challenging with current technology. For example, transmission electron microscopy (TEM) based techniques[6–10] provide high spatial resolution and high precision for strain and electric field mapping, but they only offer a two-dimensional (2D) projection of the 3D device and are destructive due to the thin-film specimen requirement. Optical[11] and X-ray based methods [12–14] cause less damage to the sample and are capable of 3D mapping, yet their spatial resolution is limited to above 40 nm. Atomic force microscopy (AFM) based[15,16] and other tip-assisted techniques[17,18] can enhance the lateral resolution but are restricted to 2D mapping of the surface or very shallow regions. A more detailed comparison is given in the Supplementary Information. Another electric field and strain detection technique based on nitrogen-vacancy centres in diamond has demonstrated high sensitivity and nanometre scale spatial resolution[19–22]. However, the diamond material limits its applicability for characterizing modern semiconductor devices with complex structures.

Here we propose and demonstrate use of single erbium ($Er^{3+}$) ions as atomic sensors for electric field and strain sensing within the channel of a semiconductor device (Fig. 1a). As illustrated in Fig. 1b-d, the optical absorption wavelength of a single ion is sensitive to its surrounding environment due to its electrostatic interaction with the electric field and crystal lattice. By fitting the spectral shifts of a single $Er^{3+}$ ion to a Hamiltonian, one can determine the local electric field and strain around the ion. As an example, the relevant simulation of an $Er^{3+}$ ion with cubic symmetry is given in the Supplementary Information. To map the electric field and strain of the device channel, a low density of Er ions can be introduced into the channel, either during or after device fabrication. (Fig. 1a). By virtue of the 10-MHz linewidth that has been reported for a single $Er^{3+}$ ion[23], individual ions can be identified using their absorption wavelengths. Furthermore, their locations can be determined with an accuracy of about 1 nm using a micrometre size magnet, as detailed below. Finally, tests on a large number of identical devices produce a statistical 3D map of the electric field and strain in this type of device. This would also allow the strain variability among the devices to be determined. Figure 1 e and f compare the precision and resolution of existing strain and field imaging techniques with our Er based approach. This highlights the high spatial and strain resolution of this approach, in addition to its 3D mapping and deep sensing capabilities.

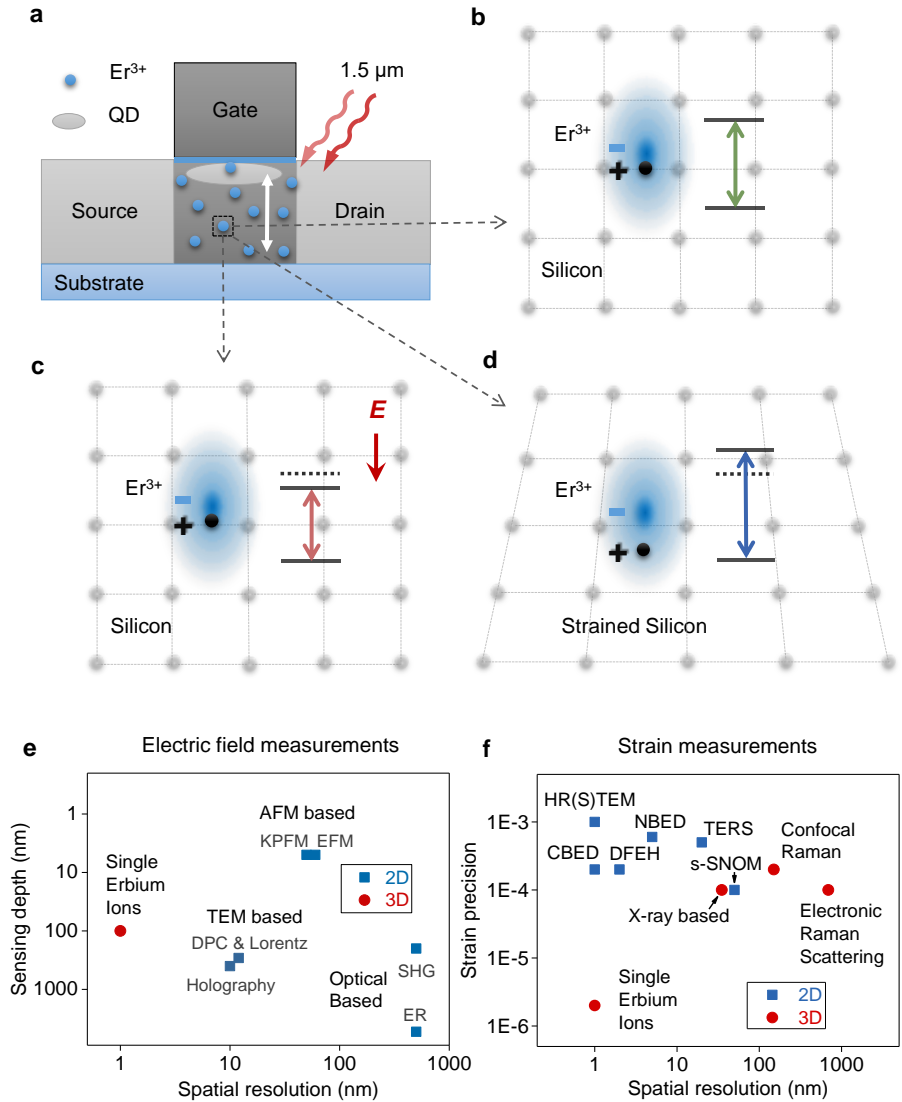

**Fig. 1 Single rare-earth ions as atomic-scale probes for electric field and strain. a**, The channel of an ultra-scaled transistor is doped with Er ions, which act as atomic scale probes of field and strain. Here the transistor works as a single electron transistor (SET) for recording the $Er^{3+}$ photo-ionisation spectra. **b,** An erbium ion in the silicon lattice forms an electric dipole. **c,** Applied electric fields will induce a Stark shift on the optical transition of the ion. **d,** Lattice strain also induces Stark shifts on the ion. This interaction can be modelled using Hamiltonians such as Fig. S1, presented in the Supplementary Information. **e,** A comparison of the spatial resolution and sensing depth between our proposed approach and other reported electric field mapping techniques. Here we define the sensing depth of our approach as the charge sensing range of the SET (indicated by the white arrow in **a**). The blue squares indicate techniques capable of only 2D mapping, while the red dots represent 3D mapping techniques. **f,** A Comparison of the spatial resolution and strain precision between our method and other reported strain detecting techniques. List of abbreviations: DPC-TEM, differential phase contrast imaging with transmission electron microscopy; EFM, electric force microscopy; KPFM, Kelvin probe force microscopy; SHG, optical second-harmonic generation; ER, electro-reflectance. CBED, convergent-beam electron diffraction; NBED, nano-beam electron diffraction; HR(S)TEM, high-resolution (scanning) TEM; DFEH, dark-field electron holography; TERS, tip enhanced Raman scattering microscopy; s-SNOM, scattering-type scanning near-field optical microscopy.

In this work, we demonstrate that a single $Er^{3+}$ ion can be employed as a sensitive probe of applied electric field and strain in an ultra-scaled silicon transistor. The single $Er^{3+}$ ion is addressed using a hybrid optical–electrical technique where electrical signal is used to detect $Er^{3+}$ ion resonant photo-ionisation. More details regarding this spectroscopic technique and device preparation can be found in Ref 24. Figure 2a shows an image of a typical fin field-effect transistor (FinFET) device used in this work. It is operated at 4.2 K in liquid helium to achieve highly sensitive charge detection. Under appropriate bias conditions, a current (I) arises from electron tunnelling via quantum dots (QDs) formed in the device channel[25], and the device operates as a single electron transistor (SET). Isolated peaks in the current-voltage (I-$V_G$) trace are observed due to Coulomb blockade[26], as illustrated in Fig. 2b. When the charge environment changes around the QD, the associated Coulomb peak will shift, and the tunnelling current through the QD will be altered. To measure the spectrum of a single $Er^{3+}$ ion, the device is illuminated with a wavelength tuneable laser. When the laser wavelength is resonant with an $Er^{3+}$ ion, it can drive the optical transition of the $Er^{3+}$ ion and this further induces ionization and current enhancement. An optical spectrum of the $Er^{3+}$ ion is collected by recording the tunnelling current as a function of wavelength as shown in Fig. 2c.

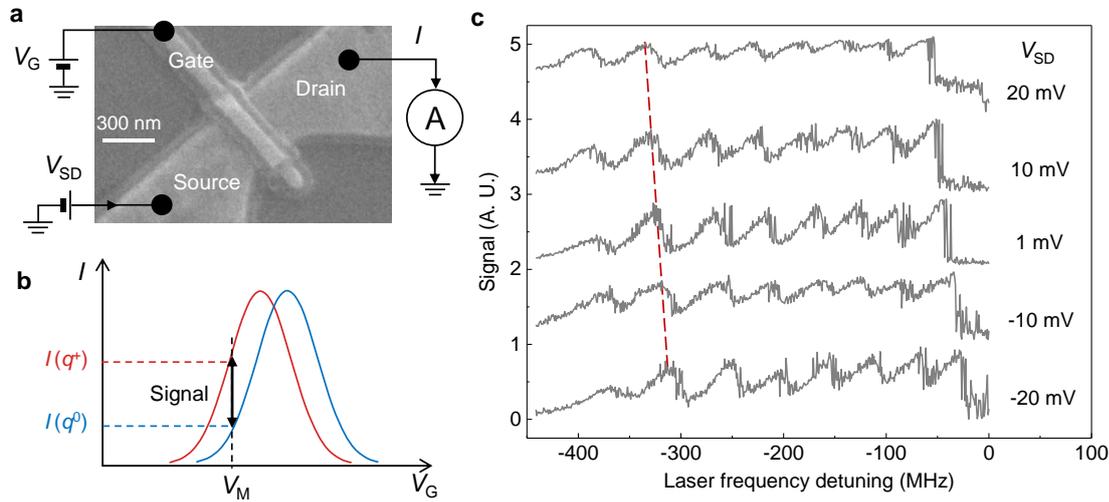

**Fig. 2 The single ion spectroscopy method. a**, A scanning electron micrograph of a typical device used in this study showing the electrical connection to the device. **b**, The SET charge-sensing scheme. Resonant laser light will ionise the $Er^{3+}$ ion and induce a transient shift of the current (I) – gate voltage ($V_G$) curve towards lower gate voltages, causing a change in current from $I(q^0)$ to $I(q^+)$. **c**, The spectra of one $^{167}Er^{3+}$ ion showing monotonic shift as the source-drain voltage ($V_{SD}$) changes. The eight peaks show optical transitions which preserve the nuclear projections of the $I=7/2$ isotope, $^{167}Er$. Since the exact shape and amplitude of the Coulomb peak varies with $V_{SD}$, these spectra are normalised for comparison.

The source-drain voltage $V_{SD}$ is first used to tune the electric field in the device. In a traditional field-effect transistor, the longitudinal electric field along the channel is proportional to source-drain voltage in the 'linear region'. In an ultra-scaled transistor, defects in the channel often lead to variation in the nominally linear relationship. Nevertheless, considering $V_{SD}$ here is on the order of 10 mV, we expect the longitudinal electric field to have an approximately linear dependence on $V_{SD}$. Figure 2c shows the spectra detected at a fixed gate voltage $V_G = 266$ mV at 5 different $V_{SD}$ values. For the data presented in Fig. 2c, a 4 T magnetic field is used to ensure the electronic Zeeman splitting of the $Er^{3+}$ ion is much greater than the ~GHz electronic-nuclear coupling, which in turn exceeds the nuclear Zeeman and Stark susceptibilities. As a result, the hyperfine sub-structure within each electronic Zeeman arm can be resolved, and the resonant peaks are narrow enough to discern the Stark shift. To meet these conditions, suitable but different magnetic fields are used in the other measurements of this work. The guideline in Fig. 2c indicates that there is a monotonic shift in the spectrum when the $V_{SD}$ varies from -20 mV to 20 mV.

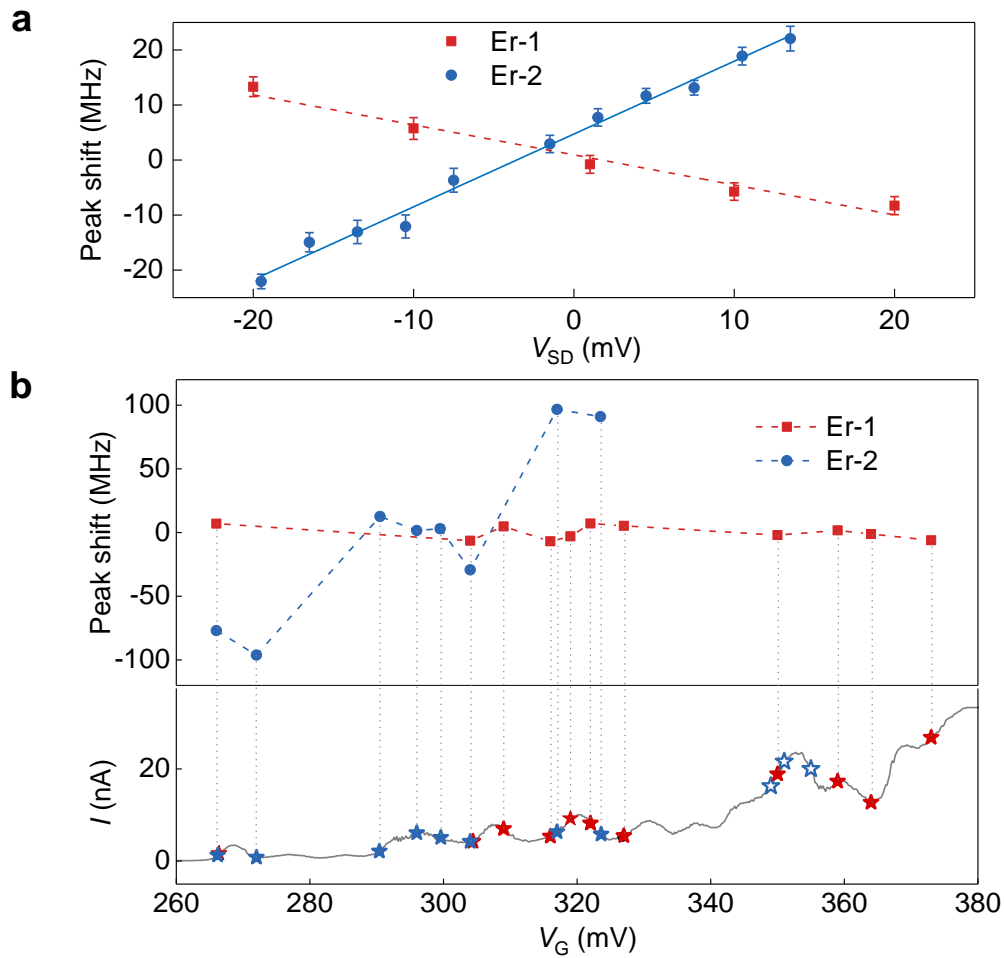

**Fig. 3 Voltage induced Stark shifts observed for two spatially separate Er ions. a**, The shift in optical transition frequencies of Er-1 (red square) and Er-2 (blue dot) as a function of $V_{SD}$. The data is overlaid with linear fits. **b**, **Top**: The Stark shift as a function of $V_G$ for each site. The error bars are smaller than the point size. **Bottom**: The $I - V_G$ curve of the transistor. Red and blue stars illustrate where the data was collected, with regard to the Coulomb peak structure. There is no Stark shift data for the three blue hollow stars around 352 mV, due to a lack of response of the photo-ionisation signal from Er-2, while signal from Er-1 could still be collected in this $V_G$ region.

The asymmetric line broadening has been previously observed and attributed to the charge fluctuation in the device[24]. To determine the peak position for different spectra, we fit each absorption peak to a bi-Gaussian function that takes the asymmetry into account. In Fig. 3a, the peak position is plotted as a function of $V_{SD}$ for two different $Er^{3+}$ ions in the same FinFET labelled as 'Er-1' and 'Er-2'. The absolute wavelength of the resonant peak is around 1541.643 nm for Er-1 and 1541.634 nm for Er-2. Both ions show a linear dependence between the peak position and the applied voltage.

We subsequently investigated the susceptibility of each $Er^{3+}$ ion to gate voltage $V_G$. Figure 3b shows that the gate voltage directly impacts the electronic configuration and the formation of QDs that are required for charge detection. For $V_G < 280$ mV, energy levels from only one or a few QDs contribute to the transport current, and discrete Coulomb peaks are observed. For $V_G > 290$ mV, more QDs start to form and contribute to the transport current, so the Coulomb peaks often overlap. For this reason, a complex relationship is observed between the electric field at each $Er^{3+}$ ion and $V_G$ (Fig. 3b). Consequently, this allowed us to study the Stark shift due to not only the global electric field from the gate, but also local fields created by QD charge reconfiguration.

The upper panel of Fig. 3b shows the Stark shift of the two $Er^{3+}$ ions measured at a series of gate voltages, and this is quite different from the Stark shift due to the source-drain voltage as previously shown in Fig. 3a. The spectral shift of Er-1 is less than 15 MHz as $V_G$ is increased from 266 to 373 mV, while Er-2 shows a ~200 MHz shift as $V_G$ is decreased from 266 to 323.5 mV. The smaller frequency shifts observed for Er-1 could be due to a large angle between its dipole moment and the applied field, or a large separation between the ion and the gate.

Conversely, the spectral response of Er-2 appears to be strongly influenced by the global electric field of the gate. Furthermore, when the gate voltage moves from one Coulomb peak to another, there is an abrupt shift of the peak frequency. This is likely caused by a local electric field change due to charge reconfiguration of the surrounding QDs. Another interesting observation is that the photo-ionisation signal from Er-2 is not present at the Coulomb peaks at $V_G = 352$ mV, while the signal from Er-1 is still clearly visible. This indicates that Er-2 is further from the charge sensing QD than Er-1, for $V_G = 352$ mV.

These observations not only confirm that the Er-related spectrum is sensitive to the applied electric field, but also indicate that the local electric field in an ultra-scaled transistor can be measured directly to optimise the design. This is particularly important as local fields cannot be readily simulated, and affect the performance of ultra-scale transistors.

Here, we employ a highly simplified model to estimate the Stark coefficient by assuming a constant electric field between the source and drain leads. This model gives a Stark coefficient of -5 kHz/V cm$^{-1}$ and 13 kHz/V cm$^{-1}$ for Er-1 and Er-2, respectively. This has not been measured for $Er^{3+}$ ions in silicon, but similar values were measured for ensembles of Er atoms in other host materials, such as 25 kHz/V cm$^{-1}$ in a crystalline $LiNbO_3$ waveguides and 15 kHz/V cm$^{-1}$ in a silicate fibre[27]. More accurate Stark effect studies could be achieved, if specially designed devices are used to create a more uniform electric field in the host material and the $Er^{3+}$ ions can be located by the Zeeman effect under a magnetic field gradient.

In addition to investigating the spectral response under applied fields, we now turn to the spectral response of single $Er^{3+}$ ions to applied strain. Much like an applied field, strain will shift the optical transitions of an $Er^{3+}$ ion due to the Stark effect. To measure the spectral response under applied strain, the sample is mounted onto a piezoelectric ceramic (PZT8) actuator with epoxy resins (Fig. 4a). Here the sample is thinned down to 80 μm thick using chemical mechanical polishing so that the applied stress can effectively reach the top layer where the ultra-scaled transistors are located. As the strain tuning voltage ($V_{str}$) was applied in the polarization direction ($z$) of the piezo-actuator, biaxial strain was generated in the transverse ($x$-$y$) plane. A calibration test with a strain gauge attached to the sample indicated a linear relationship between the strain and $V_{str}$ with a gradient of $3.432 \pm 0.011 \times 10^{-8}$ / V (See the Supplementary Information for more details). Fig. 4b shows the spectral shift of two $Er^{3+}$ ions as a function of applied strain. Both ions show a linear response in optical absorption frequency as a function of strain. The slope is constant over a range in strain of $\pm 10^{-5}$. A linear fit gives a strain coefficient of $3.8 \pm 0.1$ MHz / $10^{-6}$ for Er-3, and $5.1 \pm 0.2$ MHz / $10^{-6}$ for Er-4.

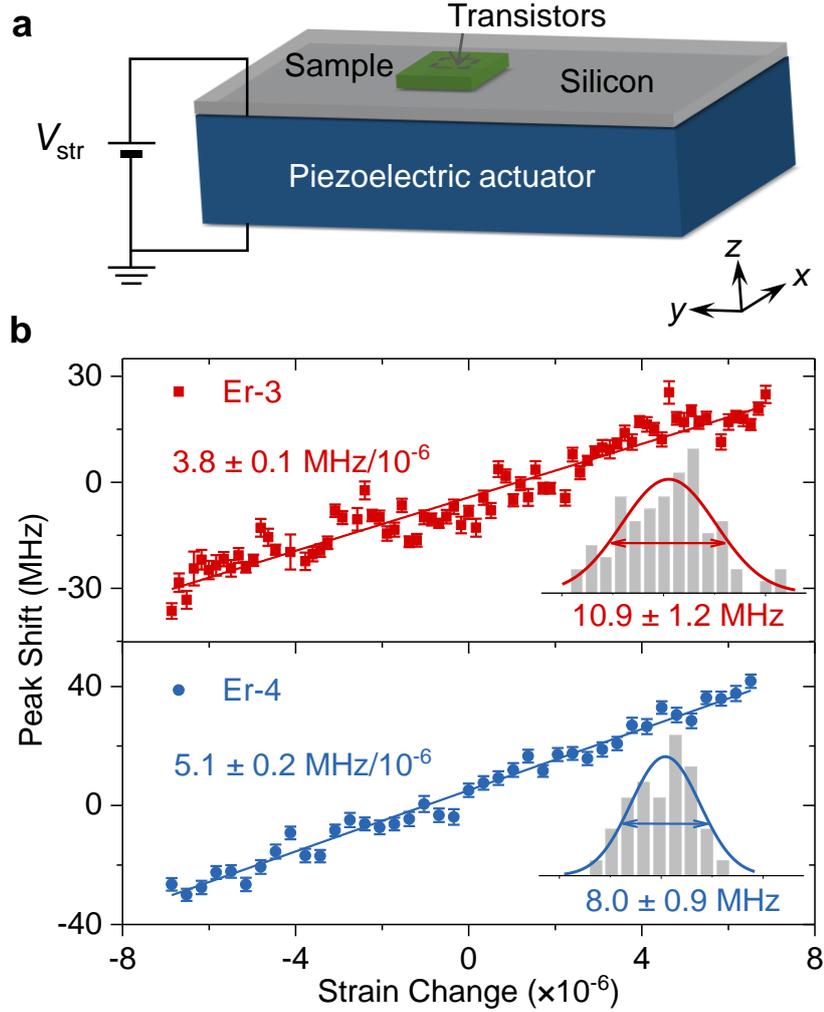

**Fig. 4 The spectral response of two Er³⁺ ions to applied strain. a,** The sample is glued to a piezoelectric actuator for strain tuning. **b,** The observed spectral shift of Er-3 and Er-4 under induced bi-axial strain in the *x-y* plane. The solid lines are linear fits. The insets show the statistics of the spectral peak deviation from the linear fit.

To estimate the measurement precision of the strain, the following 3 factors were considered. First, the fitting error of the spectral peaks was about 2 MHz for both ions, which mainly originated from rapid charge fluctuations which modulated the current signal and broadened the spectra. Secondly, the slow drift of the Er optical absorption wavelength and the hysteretic response of the piezo-actuator led to long-term fluctuations. For these reasons, we chose the measurement time of each spectrum to be 5-10 minutes. As an example, the set of spectra for generating the Er-3 data points in Fig. 4b is presented in the Supplementary Information. This resulted in a peak stability of 10.9 ±1.2 MHz for Er-3 and 8.0 ±0.9 MHz for Er-4. To obtain these values, we calculated the peak deviation against the linear fit (the solid lines in Fig. 4b), and fit the deviation to a Gaussian distribution as shown in the insets of Fig. 4b. Finally, the ratio of the peak stability to the strain coefficient gave a strain measurement precision of approximately $2.9 \times 10^{-6}$ and $1.6 \times 10^{-6}$ for Er-3 and Er-4, respectively.

The approach to electric field and strain mapping proposed in this work also requires knowledge of the location of each measured ion. This information can be obtained by measuring the effect of magnetic field gradients on each ion, following the same principles of Magnetic Resonance Imaging (MRI). Here a micro-magnetic or scanning magnetic tip[28] can be used to create such field gradients, and the optical Zeeman tensor **Δg** can be calibrated using a homogeneous vector magnetic field. For the $^4I_{13/2}$ - $^4I_{15/2}$ transitions of $Er^{3+}$ ions, the components of the **Δg** tensor (the scalar field susceptibilities) are typically on the order of 10 MHz/Gs. Given the 10 MHz optical linewidths measured in this work, a magnetic field gradient of 10 Gs/nm, generated by a Co tip at a distance of 100 nm [29], is sufficient to locate an $Er^{3+}$ ion with a spatial resolution of 1 nm.

In addition to locating the Er ions, it is critical to accurately extract the electric field and strain values from the measured spectra. This process is explained in the Supplementary Information with an $Er^{3+}$ ion with cubic symmetry. To obtain the best accuracy of the final results, one should consider the following factors. First, Er is known to form a large variety of sites in silicon[30], and large inhomogeneous broadening around several gigahertz is observed even for samples containing Er ions all having the same site symmetry[31]. Thus, the variation between different Er ions with the same site symmetry needs to be taken into consideration, and the high resolution spectra on more crystal field levels can help reduce the uncertainty. Secondly, the strain data is collected at cryogenic temperatures, but could be transformed to a room temperature strain value using the thermal dependence of the lattice constant. Thirdly, the incorporation of Er ions into silicon introduces defects that affect the strain distribution in the device channel. This is expected to play a minor role compared to the density of defects produced by the multiple processes of fabricating modern 10-nm node devices. However, to minimise this, Er ions with the minimum required fluence should be introduced into the wafer before device fabrication and a careful annealing strategy should remove excess defects.

Furthermore, this is the first observation of Stark tuning on single rare-earth ions in solids, which is useful for quantum computing applications. For example, recent progress demonstrated a long spin coherence time of $^{167}Er^{3+}$ ions in $Y_2SiO_5$[32] and the coupling of single $Er^{3+}$ ions to a silicon cavity[33]. One important element for photonic quantum circuits is rapid control over the ion-cavity coupling[34], and this can be achieved with the single-ion Stark tuning using a local electrical contact.

In conclusion, we have shown that the spectrum of a single $Er^{3+}$ ion can be used to detect electric field and strain. The single ion spectra are sensitive to not only the overall electric field but also the local electric field, and the latter proves the necessity of experimental 3D mapping of the electric field in ultra-scaled transistors. The strain measurement shows a high precision, of better than $3\times10^{-6}$. Combined with the estimated 1-nm spatial resolution, the proposed method would enable nano-scale 3D mapping of the local strain and electric field in the channel of ultra-scaled transistors.

Acknowledgements
We thank J. Bartholomew, R. Ahlefeldt, Z. Wu, S. Horvath, and M. Reid for helpful discussions. We would also like to acknowledge M. Doherty for his assistance in deriving the Hamiltonian of strained Er-Si. This work was supported by the ARC Centre of Excellence for Quantum Computation and Communication Technology (Grant No. CE170100012) and the Discovery Project (Grant No. DP150103699). Q.Z. and J.D. acknowledge support from NNSFC (Grants No. 81788101, No. 11227901), the CAS (Grants No. QYZDY-SSW-SLH004 and No. XDB01030400), the 973 Program (Grants No. 2013CB921800), and China Postdoctoral Science Foundation (Grant No. BX201700230, No. 2017M622001). C.Y. acknowledges support from a Discovery Early Career Researcher Award (Grant No. DE150100791).

Author contributions
S.R. proposed the initial idea; Q.Z., G.H., G.G.d.B., M.R., and C.Y. performed the experiments and conducted the simulations; B.C.J. and J.C.M. performed the implantation; M.J.S, C.Y., and S.R. supervised the project; all authors contributed to the data analysis and the writing of the manuscript.

Additional information
Supplementary Information is available in the online version of the paper. Reprints and permissions information is available online at www.nature.com/reprints. Correspondence and requests for materials should be addressed to C.Y.

Competing financial interests
The authors declare no competing financial interests.

# Supplementary Information for
# Single rare-earth ions as atomic-scale probes in ultra-scaled transistors

**Characterising the strain field of a single Er ion**

Fig. S1 shows the simulated spectral response of an Er ion with cubic site symmetry under different stresses. The $^4I_{15/2} \rightarrow {}^4I_{13/2}$ optical transition for Er ions in Si is completely described by the Lea-Leask-Wolf (LLW) crystal field Hamiltonian $H_{CF}$[1]. The Hamiltonian parameters for well annealed and unstrained $T_d$ Er sites in Si were determined by Przybylinska et al.[2] From first principles, it is possible to determine the perturbation to this Hamiltonian from a strain field, $\boldsymbol{\sigma} = \sum_{i,j=x,y,z} \sigma_{i,j}$, and electric field, $\boldsymbol{E} = \sum_{i=x,y,z} E_i$. By considering the character table for the $T_d$ point group, one arrives at the following Hamiltonian:

$$H = H_{CF} + A(\sigma_{xx} + \sigma_{yy} + \sigma_{zz})J^2$$
$$+ B[(2\sigma_{zz} - \sigma_{xx} - \sigma_{yy})(2J_z^2 - J^2) + (\sigma_{xx} - \sigma_{yy})(J_x^2 - J_y^2)]$$
$$+ C[\sigma_{xy}\{J_x,J_y\} + \sigma_{xz}\{J_x,J_z\} + \sigma_{yz}\{J_y,J_z\}]$$
$$+ D[(E_x + E_y)\{J_x,J_y\} + (E_x + E_z)\{J_x,J_z\} + (E_y + E_z)\{J_y,J_z\}]$$

The xyz coordinate is chosen to be consistent with Si crystalline <100> coordinate. The empirical constants in the perturbation Hamiltonian can be determined with a calibration experiment on an unstrained and well annealed Er-doped Si wafer. The constants *A*, *B* and *C* can be determined by applying a known stress.

Specifically, hydrostatic stress only introduces an isotropic strain tensor with equal diagonal terms ($\sigma_{xx} = \sigma_{yy} = \sigma_{zz}$) and zero non-diagonal terms ($\sigma_{xy} = \sigma_{xz} = \sigma_{yz} = 0$). So a constant *A* could be directly determined by a hydrostatic stress experiment shown in Fig. S1a. Uniaxial stress along the <100> direction also generates a diagonal strain tensor with non-diagonal terms equal to zero ($\sigma_{xy} = \sigma_{xz} = \sigma_{yz} = 0$), while a diagonal shear component will result in non-equal diagonal terms. Once *A is known*, *B* could be derived from a uniaxial stress experiment along <100> (Fig. S1b). Uniaxial stress along <111> will introduce a strain tensor with isotropic diagonal terms ($\sigma_{xx} = \sigma_{yy} = \sigma_{zz}$) and non-zero non-diagonal terms, from which the constant *C* could be determined.

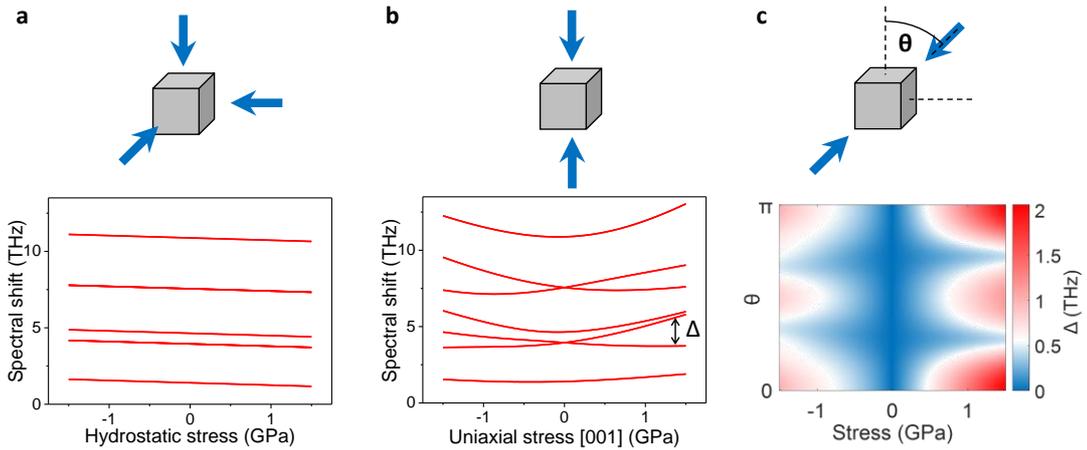

**Fig. S1 Spectral shift simulation of cubic Er site under different stresses. a,** The hydrostatic stress keeps the crystal symmetry and simply shifts all the transitions linearly. **b,** Uniaxial stress [001] breaks the crystal symmetry, introducing a non-linear dependence and splitting to the crystal levels of the Er site. **c,** The dependence of the level splitting (Δ indicated in b) on uniaxial stress magnitude and polar angle (θ) in the (010) plane.

For the simulation in Fig. S1, the crystal field Hamiltonian parameters are from results reported in Ref. 28 and 29. There is no experimental work directly measuring constants *A*, *B* and *C*, so we set them phenomenologically according to the results from Fig. 4 and spectroscopic measurements of ensemble $Er^{3+}$ ions doped in strained silicon structure[3–5], with 2*A* = *B* = 2*C* = 2 THz per unit of strain.

Similarly, the constant *D* could be determined with a known electric field. For the above Hamiltonian, it is important to note that both shear stress and electric field have the same phenomenological effect on the Er ion. Fortunately, once the physical location of the site is determined, the local electric field can be accurately determined using Finite Element Analysis (FEA).

With a *priori* knowledge of the stress Hamiltonian, the determination of the stress tensor in a microscopic voxel has already been demonstrated, using several NV centres in diamond[6]. For cubic Er sites in silicon, however, the lack of non-equivalent site orientations limits the usefulness of comparing neighbouring sites. Instead, each Er ion provides much richer spectroscopic information, given the fourteen (optically accessible) crystal field levels in the $^4I_{13/2}$ manifold. With a resolution of ~ 10 MHz on the optical transitions, it should be possible to determine very precisely the magnitude (but not direction) of the hydrostatic, diagonal and non-diagonal shear strain components by studying all 14 optically accessible levels.

In conjunction with a magnetic field rotation study for each site, some prior knowledge of the overall strain field direction should help determine some of the directional components of the local strain field. Under ideal conditions, it may even be possible to unambiguously determine all 6 independent components of the local strain tensor *σ* at the location of some sites.

**The strain calibration test**

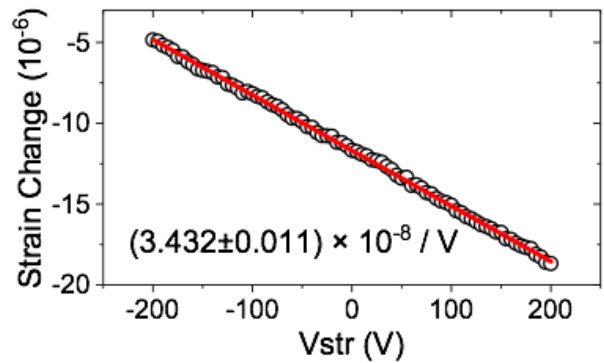

**Fig. S2** Measurements with the strain gauge reveal a linear dependence of the actuator-induced strain on *V*str. The circles are averaged data from 5 scans back and forth between *V*str=+/- 200 V.

**A set of spectra recorded in a strain tuning measurement**

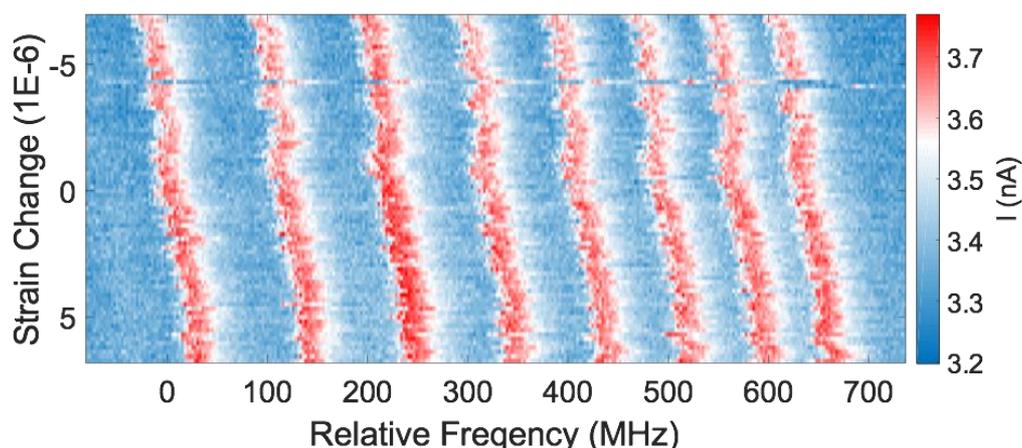

**Fig. S3** Spectra recorded for Er-3 as a function of laser frequency and the change in strain (converted from the *V*str value). The colour scale indicates the photo-ionisation current. The abrupt spectral shift at a strain change of -4.3 x $10^{-6}$ (*V*str = 125 V), is due to a temporary unlock of the cavity-stabilised laser.

**Nano-scale strain mapping in a semiconductor device.**

Strain engineering could introduce band bending to lower the effective mass of carriers, and enhance their mobility. Thorough knowledge of the strain distribution will provide valuable information on the manufacturing process and provide valuable input parameters for device modelling. There are different ways of measuring strain at different scales[7] but measurements on the transistor level are definitely the keystone. Below is a comparison between different techniques used for strain characterization.

| Methods | X-Ray based | TEM based | s-SNOM & TERS | Confocal Raman | Single Er ions |
|---|---|---|---|---|---|
| Precision | $10^{-3}$ to $10^{-4}$ | $10^{-4}$ | $10^{-3}$ to $10^{-4}$ | $10^{-3}$ to $10^{-4}$ | ~$10^{-6}$ |
| Spatial resolution | ~40 nm | 2 nm | 20 - 50 nm | ~500 nm | ~nm |
| Image dimension | 3D | 2D | 2D | quasi-3D | 3D |
| Non-destructive? | √ | × | √ | √ | √ |

Non-destructive strain information can be obtained with X-ray diffraction[8–11]. However, its spatial resolution is typically tens of nanometres if huge synchrotron facilities are employed[12]. Recent results show that X-ray ptychography reaches a high lateral resolution of 14.6 nm, which could be applied to strain mapping[13]. However, complex simulations and reference samples are needed.

There are several strain sensing methods[14,15] based on transmission electron microscopy (TEM). Overall, they have the best spatial resolution and accuracy. But dedicated sample preparation is necessary for this technique. Specimens need to be milled to electron transparency thicknesses of 100-300 nm. This procedure is time-consuming and likely to cause strain relaxation. Again, complex simulations and reference samples are required.

Confocal Raman based techniques are more suitable to strain analysis on a larger scale. Generally, its spatial resolution is not enough for transistor level mapping. Furthermore, cryogenic temperatures are needed for $10^{-4}$ order accuracy[16]. Spatial resolution could be improved with scattering-type scanning near-field optical microscopy[17] (s-SNOM) or tip enhance Raman spectroscopy[18] (TERS). But, they are limited to shallow subsurface detection because of the near-field penetration depth (~20 nm).

Compared to the above approaches, the outstanding advantages of the method described here are the combination of high spatial resolution, 3D mapping and non-destructivity. This method introduces only a small perturbation to the specimen, including illumination with a 1550 nm laser well below silicon band gap energy, and $Er^{3+}$ ion implantation with fluences much lower than that normally employed by the semiconductor industry. No destructive sample treatments like milling are required. As silicon is transparent to 1550 nm light, this method could access regions much deeper below the surface, which is critical to strain metrology in further multi-layered integrated circuits. Another point that should be noted is that Er ions are relatively large compared to silicon. This will induce some perturbation to the surrounding silicon lattice, however, the perturbation volume ~$(0.3 \text{ nm})^3$ is much more local than the ~$(\text{nm})^3$ scale we are probing, and could also be evaluated from the Er site symmetry study mentioned above. A database of the parameters for different Er sites would be very helpful. Building up such a database will need quite a number of tests, such as a study on the spectral response of Er to strain from sites with various symmetries. Once the database is well formed, interpretation from spectra to strain would be simple.